\documentclass[12pt,preprint]{aastex}
\begin{document}

\title{\bf Lyman-$\alpha$ Absorption from Heliosheath Neutrals}

\author{Brian E. Wood\altaffilmark{1}, Vladislav V.
  Izmodenov\altaffilmark{2,3}, Jeffrey L. Linsky\altaffilmark{1},
  Yury G. Malama\altaffilmark{3}}

\altaffiltext{1}{JILA, University of Colorado, 440 UCB, Boulder, CO
  80309-0440; woodb@origins.colorado.edu, jlinsky@jila.colorado.edu.}
\altaffiltext{2}{Lomonosov Moscow State University, Dept. of
  Aeromechanics and Gas Dynamics, Moscow 119899, Russia; izmod@ipmnet.ru.}
\altaffiltext{3}{Space Research Institute (IKI) RAS, and Institute for
  Problems in Mechanics RAS, Prospect Vernadskogo 101-1, Moscow 119526,
  Russia.}

\begin{abstract}

     We assess what information HST observations of stellar Ly$\alpha$
lines can provide on the heliosheath, the region of the heliosphere
between the termination shock and heliopause.  To search for evidence
of heliosheath absorption, we conduct a systematic inspection of stellar
Ly$\alpha$ lines reconstructed after correcting for ISM absorption (and
heliospheric/astrospheric absorption, if present).  Most of the stellar
lines are well centered on the stellar radial velocity, as expected, but
the three lines of sight with the most downwind orientations relative to
the ISM flow ($\chi^{1}$~Ori, HD~28205, and HD~28568) have significantly
blueshifted Ly$\alpha$ lines.  Since it is in downwind directions where
heliosheath absorption should be strongest, the blueshifts are almost
certainly caused by previously undetected heliosheath absorption.
We make an initial comparison between the heliosheath absorption and the
predictions of a pair of heliospheric models.  A model with a complex
multi-component treatment of plasma within the heliosphere predicts less
absorption than a model with a simple single-fluid treatment, which leads
to better agreement with the data.  Finally, we find that nonplanetary
energetic neutral atom (ENA) fluxes measured by the ASPERA-3 instrument
on board {\em Mars Express}, which have been interpreted as being from the
heliosheath, are probably too high to be consistent with the
relative lack of heliosheath absorption seen by HST.  This would argue for
a local interplanetary source for these ENAs instead of a heliosheath
source.

\end{abstract}


\section{INTRODUCTION}

     Ultraviolet observations of nearby stars from the {\em Hubble Space
Telescope} (HST) have proven unexpectedly to be very useful for studying
the outermost heliosphere.  High resolution spectra of the H~I Ly$\alpha$
line at 1215.67~\AA\ sometimes show signatures of absorption from
H~I in the heliosphere, in cases where the interstellar absorption is weak
enough to allow the heliospheric signal to be detectable
\citep{bew05b}.  Absorption is also sometimes observed from the
astrosphere of the observed star, which has provided the first
detections and measurements of winds from solar-like stars \citep{bew05a}.

     Figure~1 shows a picture of the basic heliospheric structure, which
is characterized by three sharp boundaries \citep[see, e.g.,][]{gpz99}.
The innermost is the termination shock, where the solar wind is decelerated
to subsonic speeds.  Then there is the heliopause, which separates the
plasma flows of the fully ionized solar wind and partially ionized ISM.
Although the charged constituents of the ISM cannot penetrate the
heliopause, the neutral atoms in the ISM flow can.  The
outermost boundary is the bow shock, where the ISM flow is decelerated to
subsonic speeds \citep[see, e.g.,][]{gpz99}.  The {\em Voyager~1} satellite,
which is traveling close to the upwind direction of the ISM flow seen by
the Sun, has recently crossed the termination shock at a distance of
94~AU from the Sun \citep{ecs05}, roughly consistent with the
model predictions in Figure~1.

     For most lines of sight, the Ly$\alpha$ absorption from heliospheric
H~I will be dominated by the so-called ``hydrogen wall'' region between
the heliopause and bow shock (Region~3 in Fig.~1), where models predict
that interstellar H~I will be heated, compressed, and decelerated after
crossing the bow shock \citep{vbb91,vbb93,vbb95,gpz96}.
There are currently a total of 8 detections of
heliospheric absorption.  All but one are in upwind directions.  To be
more precise, if we define $\theta$ to be the angle between the observed
line of sight and the upwind direction of the ISM flow vector seen by
the Sun, all but one of the lines of sight with detected heliospheric
absorption have $\theta < 75^{\circ}$ \citep{bew05b}.  This is
consistent with model predictions, which demonstrate that the deceleration
of H~I within the hydrogen wall relative to the ISM flow is strongest in
upwind directions.  This helps to separate the heliospheric absorption
from the ISM absorption, making the heliospheric signal easier to detect.

     Neutrals within the heliosphere interact with other particles
almost entirely via charge exchange, densities being too low for
thermal particle collisions to be important.  The population of
hydrogen wall H~I that dominates Ly$\alpha$ absorption in upwind
directions is formed by charge exchange between interstellar H~I that has
passed unimpeded through the bow shock and ISM protons that have been
heated, decelerated, and compressed at the bow shock.  However, interstellar
neutrals can also penetrate the heliopause and charge exchange with solar
wind protons in the heliosheath region in between the termination shock
and heliopause (Region~2 in Fig.~1).  Downwind lines of sight that look
through the tail of the heliosphere will have extended path lengths
through the heliosheath.  The Ly$\alpha$ absorption in such directions,
if it can be detected, will potentially be dominated by the heliosheath
rather than the hydrogen wall.

     In this paper, we focus on observations and model predictions
concerning Ly$\alpha$ absorption from heliosheath neutrals.  There are
several reasons why this is useful.  One is simply that heliosheath
absorption probes a different region of the heliosphere than the hydrogen
wall absorption that dominates nearly all previous detections of
heliospheric and astrospheric Ly$\alpha$ absorption.  Another reason is
that hydrodynamic models of the heliosphere that use a fully kinetic
treatment of the neutrals seem to predict a disturbingly large amount of
heliosheath absorption in downwind directions, which may be in conflict
with nondetections of absorption in those directions \citep{bew00,vvi02}.
We will take a closer look at these
downwind HST data to see if any evidence for heliosheath absorption can
be found.  If not, we will derive upper limits for the amount of
absorption that can be present.

     Finally, observed heliosheath absorption can potentially
be used to constrain the flux of energetic neutral atoms (ENAs) that
should be observable within our solar system.  Local ENA fluxes should be
dominated by neutrals created by charge exchange within the heliosheath
that are then directed through the termination shock towards the Sun.
Thus, locally observed ENAs basically sample the same particle population
as the heliosheath absorption.  In 2008, the {\em Interstellar Boundary
Explorer} (IBEX) will be launched, which is entirely devoted to studying
ENAs \citep{dm05}.  Constraints on ENA fluxes from Ly$\alpha$
absorption will be valuable in preparation for this mission.  Furthermore,
the ASPERA-3 instrument on board the {\em Mars Express} satellite has
already reported a detection of heliospheric ENAs \citep{ag06}.
We will make an initial assessment as to whether the ENA fluxes
observed by ASPERA-3 are consistent with heliosheath absorption
detections and/or upper limits.

\section{SEARCHING FOR EVIDENCE OF HELIOSHEATH ABSORPTION}

\subsection{A Sirius Example of Heliosheath Absorption}

     Currently the only detection of heliospheric Ly$\alpha$ absorption
in a downwind direction, where heliosheath absorption should dominate
the hydrogen wall absorption, is for the Sirius line of sight at
$\theta=139^{\circ}$.  The Sirius data are shown in Figure~2.  The
ISM Ly$\alpha$ absorption for this line of sight is derived with
the assistance of constraints provided by other ISM absorption lines
\citep{pb95b}.  There is excess absorption on both sides of
the Ly$\alpha$ line that cannot be accounted for by the ISM.
The excess on the blue side of the line (at $<-10$ km~s$^{-1}$ in
the figure) is believed to be from the stellar wind of Sirius, which is
a hot A0~V star \citep{pb95a,gh99}.  The
excess on the red side of the line (at $>40$ km~s$^{-1}$), based on
the ISM parameters derived by \citet{pb95b}, is claimed
by \citet{vvi99} to be heliospheric absorption.

     There are serious problems with this Sirius data, and with the
detection of heliospheric absorption.  The spectral
resolution is lower than any other spectrum with a claimed heliospheric
absorption detection.  Thus, the wavelength calibration will be
more uncertain.  The signal-to-noise (S/N) of the data is also low,
as Figure~2 clearly shows.  The presence of the wind absorption
complicates the analysis of the blue side of the line, including the
D~I absorption \citep{gh99}.  Finally, observations of narrower ISM
lines demonstrate that there are actually two distinct ISM velocity
components for this line of sight, which further complicates the
Ly$\alpha$ analysis.  \citet{gh99} are able to fit HST Ly$\alpha$ spectra
of both Sirius and its companion Sirius~B without any need for the
existence of excess red-side absorption, but this interpretation of the
data ideally requires the assumption of a very low D/H ratio for
the weaker of the two ISM components seen towards Sirius,
which would contradict recent claims that D/H within the Local Bubble
is constant, with a value around ${\rm D/H}\approx 1.5\times 10^{-5}$
\citep{jll98,hwm02,bew04}.  Thus, we still consider the presence of
excess red-side absorption, possibly heliospheric in origin, to be a
plausible interpretation of the data.  Despite its problems, the line of
sight does have the advantage of having one of the lowest ISM H~I column
densities ever observed, due mostly to its very short distance of
$d=2.6$~pc.  The \citet{pb95b} analysis suggests an ISM column density
(in cm$^{-2}$) of $\log N_{\rm H}=17.53$, while \citet{gh99} infer
$\log N_{\rm H}=17.81$.  Both values are lower than any other HST-observed
downwind line of sight.  With the ISM absorption being therefore weaker
than any other downwind line of sight, this would explain why
Sirius is so far the only such line of sight with a heliospheric
absorption detection.

     Figure~2 also shows the absorption predicted by a hydrodynamic
model of the heliosphere.  The model used here is of the type described
by \citet{vbb93,vbb95}, which includes a fully kinetic
treatment of the neutrals within the heliosphere.  The neutral
hydrogen density, proton density, velocity, and temperature of the
undisturbed ISM assumed in this model are: 
$n_\infty({\rm H~I})=0.18$~cm$^{-3}$,
$n_\infty({\rm H^{+}})=0.06$~cm$^{-3}$, $V_{\infty}=26.4$ km~s$^{-1}$,
and $T_{\infty}=6400$~K.  The top panel of Figure~2 shows absorption
predicted for both the upwind 36~Oph line of sight (at $\theta=12^{\circ}$)
and the downwind Sirius line of sight ($\theta=139^{\circ}$).  We choose
the 36~Oph line of sight for comparison because it represents the most
upwind detection of heliospheric absorption, while Sirius is the
only downwind detection \citep{bew05b}.
Both absorption profiles are significantly redshifted relative to the
local ISM flow speed in those directions, based on the Local Interstellar
Cloud vector \citep{rl92,rl95}.
This is why heliospheric absorption is always found on the red side
of the ISM absorption.  Since astrospheric absorption is observed from
outside the astrosphere rather than inside, it is conversely found on
the {\em blue} side of the ISM absorption.

     The absorption predicted for the Sirius line of sight is
added to the ISM absorption and the result is compared with the data
in Figure~2.  The model appears to predict too much absorption at
$>60$ km~s$^{-1}$.  Similar results have previously been found when
comparing other kinetic models with downwind observations
\citep{bew00,vvi02}.  Since charge exchange processes drive the
neutrals in the heliosphere out of thermal equilibrium, models with a
fully kinetic treatment of the neutrals would seem to be necessary to
provide the most precise description of neutral velocity distributions
within the heliosphere, which in turn should provide the most accurate
Ly$\alpha$ absorption profiles.  Nevertheless, these kinetic
models generally seem to overpredict heliosheath absorption in downwind
directions.

     However, the nature of the heliosheath absorption predicted
by the models is such that it is not easy to entirely rule out its
presence for observed downwind lines of sight.  The top panel of
Figure~2 shows that the heliosheath absorption towards Sirius is
characterized by a very broad, extended wing on the red side
of the profile.  This is very different from the absorption
profile for the upwind 36~Oph line of sight, which is dominated
by the hydrogen wall.  Heliosheath neutrals are generally much hotter
than hydrogen wall neutrals, but heliosheath neutral column densities
are much lower.  This is why the heliosheath absorption tends to be
broad but shallow compared to the hydrogen wall absorption.  Note that
there is a weak extended wing for the 36~Oph absorption that would
correspond to heliosheath absorption in that direction, but the
absorption is too weak to be detectable.  The weakness of the
wing is simply because the heliosheath is much narrower in upwind
directions (see Fig.~1).  The hydrogen wall absorption produces
a rather sharp, well-defined absorption edge, which is something that
can be compared with the data in a reasonably definitive manner to see
whether the model is consistent with the data.  However, the
broad extended wing of the heliosheath absorption in downwind directions
complicates matters, for reasons we now describe.

     One of the primary sources of systematic uncertainty in any of
these Ly$\alpha$ analyses is the reconstruction of the background
stellar Ly$\alpha$ profile \citep{bew05b}.  No reasonable change
to an assumed stellar profile will greatly change the velocity where the
Ly$\alpha$ absorption predicted by a model becomes saturated.  This is
why the sharp absorption edge provided by the hydrogen wall absorption
provides a much better diagnostic to compare with the data than the
broad absorption wing that characterizes the heliosheath absorption,
which results in the apparent disagreement with the data in
Figure~2.  The red-wing absorption discrepancy in Figure~2 could in
principle be resolved by increasing by $\sim 30$\% the background Sirius
Ly$\alpha$ fluxes from $60-120$ km~s$^{-1}$.  Other examples of how
stellar Ly$\alpha$ profiles can be modified to allow such broad wing
absorption are presented by \citet{bew00} and \citet{vvi02}.  There
is still a limit, however, to how much one
can modify the stellar profile in this fashion before the resulting
profile becomes implausible, and there is therefore a limit to
how much extended wing absorption can be present.  For example, modifying
the Sirius Ly$\alpha$ background as described above would create a
dubious asymmetry in the stellar Ly$\alpha$ absorption profile.
Quantifying exactly when ``dubious'' becomes ``completely unreasonable''
is not easy, but that is the task that we now undertake.

\subsection{Stellar Ly$\alpha$ Line Bisector Analysis}

     Most HST-observed nearby stars that are appropriate targets
to search for heliospheric and/or astrospheric
absorption are cool main sequence stars like the Sun \citep{bew05b}.
The Ly$\alpha$ lines of such stars are relatively narrow, isolated
emission lines.  Note that Sirius is a hot A0~V star where the stellar
profile is instead a very broad absorption line, so Sirius is
unfortunately not a representative example at all.  As shown in
Figure~2, heliosheath absorption will contaminate only the red side of a
stellar emission line.  Therefore, if there is unrecognized and
undetected heliosheath absorption, the stellar Ly$\alpha$ emission
profile reconstructed from the HST data will end up blueshifted with
respect to the stellar rest frame.

     The wings of the Ly$\alpha$ line will be formed in the chromosphere
of the observed star.  Although emission lines formed at higher
temperatures in the transition region can be shifted from the stellar
rest frame, the wings of Ly$\alpha$ should be centered on the star
\citep{tra88,bew97}.  We can search
for previously undetected absorption from the heliosheath by searching
for cases in which reconstructed Ly$\alpha$ profiles end up blueshifted
with respect to the star.  The requirement that a stellar emission line
be centered on the stellar radial velocity also provides a way to
limit modifications to the stellar profile, thereby yielding upper
limits to the amount of broad heliosheath absorption that can be present
in the red wing of the Ly$\alpha$ line.

     \citet{bew05b} provide a complete list of HST Ly$\alpha$
spectra of nearby stars, all 62 of which have been analyzed to measure
ISM H~I and D~I column densities, search for heliospheric/astrospheric
absorption, measure chromospheric Ly$\alpha$ line fluxes corrected
for ISM absorption, etc.  From this list we choose a selection of
stars to look for shifts in the reconstructed Ly$\alpha$ line relative
to the stellar radial velocity.  We exclude hot stars, which have
Ly$\alpha$ in absorption rather than emission.  We exclude giant stars,
which have Ly$\alpha$ emission lines that are too broad to be effective
for our purposes.  We also exclude short-period, unresolved binaries,
partly because the stellar rest frame can change due to orbital motion
during the course of the observation, and partly because in some cases
both stars contribute to the Ly$\alpha$ emission, meaning there is no
single stellar rest frame.  We only consider data with reasonably high
S/N, and with a spectral resolution of at least
$\lambda/\Delta\lambda=40,000$ to maximize our confidence in the
wavelength calibration.

     And finally, we only consider lines of sight with ISM column
densities of $\log N({\rm H~I})<18.3$.  High ISM columns potentially
degrade the effectiveness of our analysis in a couple ways.  Broad ISM
absorption limits us to the far wings of the Ly$\alpha$ profile in our
search for Ly$\alpha$ line shifts.  Also, the damping wings of the ISM
Ly$\alpha$ absorption are much stronger at higher columns and when the
absorption is not exactly centered on the rest frame of the star,
uncertainties in the ISM H~I column density can potentially lead to
a shift in the reconstructed stellar Ly$\alpha$ line that could
mimic the shifts we are looking for due to heliosheath absorption.

     Our selection criteria reduce the original sample of Ly$\alpha$
spectra to 28 targets.  These are listed in Table~1 in order of
increasing $\theta$.  Our analysis requires accurate knowledge of
the stellar photospheric radial velocity, $V_{rad}$.  The fourth column
of Table~1 lists $V_{rad}$ for all of our stars, and the last column
provides the references for these values.  The quoted uncertainties are
generally from these references, but we set a floor of 0.1 km~s$^{-1}$ for
the uncertainties in Table~1.  In cases where the observed star is a member
of a binary system (particularly $\alpha$~Cen~B and $\chi^{1}$~Ori), we
have to consider the effects of orbital motion to establish the precise
radial velocity at the time of observation, so we have increased the
quoted uncertainty accordingly.

     Figure~3 illustrates how we measure the velocities of the stellar
Ly$\alpha$ lines, using eight of the stars in Table~1 as examples.  The
figure shows the observed Ly$\alpha$ spectra and the reconstructed
stellar line profiles.  We refer the reader to \citet{bew05b} and
references therein for details on the reconstruction process, and for
complete illustrations of the line profiles.  We note that all the
stellar profiles used here yield D/H values consistent with the
${\rm D/H}=(1.56\pm 0.04)\times 10^{-5}$ value now believed to apply
throughout the Local Bubble \citep{bew04}.  Many of the analyses actually
{\em assumed} this value \citep[e.g.,][]{bew05b}, in which case the
H~I column density is directly inferred from the more easily measured
D~I column.  Since the wings of the H~I absorption depend solely on
the H~I column density and not on the doppler parameter, the wings of
the stellar Ly$\alpha$ profile can then be inferred simply by
computing the Ly$\alpha$ opacity profile, $\tau_{\lambda}$, from the H~I
column density and multiplying the observed wings by $\exp{\tau_{\lambda}}$.

     In Figure~3, we focus only on the lower flux part of the Ly$\alpha$
profile where we can measure meaningful line
centroids for comparison with the radial velocity.  The top parts
of the profile are not useful for our purposes for two reasons.
The first is that the section of the profile above the saturated core of
the ISM H~I absorption is in reality completely unconstrained by the
data.  The second is that cool star Ly$\alpha$ line profiles (including
that of the Sun) can have asymmetric self-reversals near line center
\citep[see][]{bew05b}, meaning that this part of the profile will
not necessarily be centered on the star.

     The horizontal dotted lines in Figure~3 indicate flux levels where we
measure the centroid of the line.  Collectively these centroids form a
line bisector.  These bisectors are shown as near-vertical solid lines in
the figure.  For five of the stars (70~Oph~A, $\epsilon$~Ind, $\zeta$~Dor,
$\tau$~Cet, and $\epsilon$~Eri) the bisector ends up where we expect it,
very close to the stellar radial velocity.  However, for the three most
downwind lines of sight ($\chi^{1}$~Ori, HD~28205, and HD~28568) the
bisector is significantly blueshifted with respect to the star.  This is
exactly the signature of undetected heliosheath absorption that we are
looking for, and the downwind direction is where we expect to see it.

     We compute the average velocity of the bisector, $V_{bis}$, for
each star, ignoring the lowest two flux levels where low S/N in the
extreme far wings of the line sometimes causes unreasonable excursions
in the bisector (e.g., HD~28568 in Fig.~3).  These bisector velocities
are listed in Table~1.  We define the discrepancy between these
velocities and the stellar rest frame as $\Delta V\equiv V_{bis}-V_{rad}$.
In Figure~4, $\Delta V$ is plotted versus $\theta$ for all the stars
listed in Table~1.

     For the three $\theta>160^{\circ}$ cases the Ly$\alpha$
profile is significantly blueshifted.  This blueshift is
common to all $\theta>160^{\circ}$ lines of sight, is not seen in
other directions, and is exactly in the direction where we would
expect heliosheath absorption to be maximized.  Thus, we conclude
that it is highly likely that heliosheath absorption is responsible
for these Ly$\alpha$ blueshifts.  For each of these stars, we can
construct a stellar profile that is forced to be centered on the stellar
rest frame by reflecting the blue wing of the original reconstructed
profile across the stellar radial velocity onto the red wing.  The
flux difference between the red wing of this profile and the original
one provides an estimate of the heliosheath absorption that is present,
which can be compared with the predictions of hydrodynamic models of
the heliosphere.

     Other than the three $\theta>160^{\circ}$ sight lines, the
reconstructed Ly$\alpha$ profiles are encouragingly close to their
expected locations at the stellar radial velocity.  The average and
standard deviation of the $\theta<160^{\circ}$ data points is
$\Delta V=0.0\pm 1.4$ km~s$^{-1}$.  The scatter can be easily explained
by uncertainties in the spectral wavelength calibration, and systematic
errors present in the profile reconstruction
process.  We note that most of the HST data for the stars listed in
Table~1 are STIS/E140M spectra, which have demonstrated 1$\sigma$
random wavelength calibration uncertainties of $\pm 0.9$ km~s$^{-1}$
\citep{bew05b}.  This same analysis also found a systematic offset of
$-1.2$ km~s$^{-1}$ for STIS/E140M Ly$\alpha$ spectra.  We corrected
all the STIS/E140M spectra for this shift before starting the bisector
analysis.  The centering of the $\Delta V$ values
on 0 km~s$^{-1}$ (for $\theta<160^{\circ}$) implies that
this correction successfully removed all significant systematic
wavelength calibration problems.

     None of the $\theta<160^{\circ}$ profiles is shifted away from the
stellar rest frame by more than 3 km~s$^{-1}$.
These results imply that a reconstructed stellar Ly$\alpha$ line should
be within 3 km~s$^{-1}$ of the stellar rest frame to be reasonable.  This
limits how much the Ly$\alpha$ profiles can be changed to allow for
potential heliosheath absorption.  We can therefore use these limits to
establish upper bounds on the amount of heliosheath absorption that can
be present for the $\theta<160^{\circ}$ lines of sight listed in Table~1.

     To construct an upper limit to the red wing of the Ly$\alpha$ profile,
we define an upper limit bisector that at all flux levels is
$+3$ km~s$^{-1}$ from the stellar radial velocity or $+3$ km~s$^{-1}$ from
the velocity of the original bisector, whichever is largest.  An upper
limit for the red wing of the Ly$\alpha$ profile is then derived by
reflecting the blue wing of the original reconstructed profile across the
upper limit bisector.  Figure~5 shows the result for the case of
$\tau$~Cet.  The difference between the upper limit profile and the
original one represents an estimate of the maximum amount of heliosheath
absorption that can be present in that direction.

     One final issue worthy of mention concerns astrospheric absorption.
If we observed a star through the downwind tail of its astrosphere, it
is possible that astrosheath absorption would lead to a reconstructed
stellar Ly$\alpha$ profile that is {\em redshifted} relative to the
star rather than blueshifted, keeping in mind that astrospheric
absorption appears on the opposite side of the Ly$\alpha$ line from the
heliospheric absorption (see \S2.1).  However, Figure~4 shows no data
points with $\Delta V>+3$ km~s$^{-1}$.  Given that a heliospheric effect
is only seen for $\theta>160^{\circ}$, we conclude that either none of
these 28 sight lines has $\theta>160^{\circ}$ for the stellar astrosphere,
or if there are any $\theta>160^{\circ}$ astrospheric lines of sight the
ISM surrounding the star contains no neutrals to provide Ly$\alpha$
absorption, which should be a common occurrence within the mostly ionized
Local Bubble \citep{bew05b}.

\section{MODELS WITH SINGLE- AND MULTI-COMPONENT PLASMAS}

     As an example of how the heliosheath absorption detections and
upper limits can be used, we now compare the Ly$\alpha$ data with the
predictions of a pair of heliospheric models.  One of the models is
the kinetic model described in \S2.1 and used in Figure~2.  This model
is of the \citet{vbb93,vbb95} type, which uses a fully
kinetic treatment of the neutrals but treats the plasma as a single
hydrodynamic fluid.  This plasma treatment is simplistic.  The
best evidence for this concerns pick-up ions inside the termination shock
(Region~1 in Fig.~1).  Pick-up ions originate from charge exchange
between outflowing solar wind protons and inflowing ISM neutrals, which
creates a solar wind neutral H atom and an inflowing ion that is then
``picked-up'' by the solar wind.  Observations show that the pick-up
ions are not thermalized with the solar wind, and the pickup ions also
show non-Maxellian velocity distributions \citep[e.g.,][]{gg04}.
Thus, combining solar wind plasma and pick-up ions into a single fluid
in a heliospheric model, as is typically done, will lead to inaccuracies
in the particle distributions.

     \citet{ygm06} have recently
attempted to solve this problem by modifying the Baranov \& Malama
heliospheric modeling code to allow a complex multi-component treatment
of the plasma, as well as including the usual full kinetic treatment of
the neutrals.  We consider here a multi-component model of this type
using the same input parameters assumed in the single-component model
(see \S2.1).  Figure~1 shows the differences in global heliospheric
structure suggested by the two models.

     Figure~6 compares the absorption predicted by the single-
and multi-component models with the Ly$\alpha$ absorption observed
for four downwind lines of sight, including the Sirius line of sight
discussed in \S2.1.  The figure focuses on the red side of the
absorption where the heliospheric absorption should be located.
Using the original reconstructed stellar profiles, the ISM absorption
alone is illustrated, and the absorption predicted by the heliospheric
models is shown after being added to the ISM absorption.

     Dashed lines in Figure~6 indicate the extra heliosheath absorption
that is allowable by the data based on the upper limit stellar profiles
derived as described in the previous section (see Fig.~5).  Models can
be considered to be consistent with the data if they do not predict
more absorption than these dashed lines, so the dot-dashed lines should
lie above the dashed lines in Figure~6.  The dashed lines have a lower
velocity limit in Figure~6.  This is because the line bisector analysis
described above is not applicable close to line center (for reasons
discussed in \S2.2), so we cannot really quantify an upper limit to the
profile near line center.  No absorption limits have been derived for
Sirius since it is a hot star where the stellar profile is a broad
absorption line rather than a comparatively narrow emission line
(see \S2.2).  However, as described in \S2.1, this is the one downwind
line of sight with a previous detection of heliosheath absorption, with
the ISM absorption underpredicting the observed absorption in
Figures~2 and 6.

     As described in \S2.2, the Ly$\alpha$ bisector analysis suggests
that $\chi^{1}$~Ori and two other very downwind lines of sight (HD~28205,
HD~28568) show evidence for broad heliosheath absorption.  The thick
solid line in the $\chi^{1}$~Ori panel of Figure~6 corresponds to the
amount of excess absorption that {\em should} be present to allow a
stellar line profile that is centered on the rest frame of the star.
For $\chi^{1}$~Ori, therefore, the model absorption in Figure~6 is best
compared with the thick solid line rather than with the data.
We could have added panels for HD~28205 and HD~28568 to Figure~6, but
this would be somewhat redundant since the $\chi^{1}$~Ori, HD~28205,
and HD~28568 lines of sight are very similar, and the $\chi^{1}$~Ori
data are in any case far superior to the others in terms of S/N.
The negative radial velocity of $\chi^{1}$~Ori profile also makes that
line of sight superior to HD~28205 and HD~28568.  A blueshifted profile
means that the bisector method can detect and quantify heliosheath
absorption at lower heliocentric velocities than for redshifted
profiles like those of HD~28205 and HD~28568, since you cannot do the
search close to the center of the stellar profile.  For $\chi^{1}$~Ori,
we can extend the dashed and solid lines down to $\sim 80$ km~s$^{-1}$
(see Fig.~6), but for HD~28205 and HD 28568 this could only be done
down to $\sim 130$ km~s$^{-1}$.

     The absorption limits (dashed lines in Fig.~6) allow a significant
amount of broad heliosheath absorption to exist while still being
consistent with the data.  Thus, the initial impression of disagreement
between the models and the data in Figures 2 and 6 is partly illusory.
The multi-component model predicts less absorption than the
single-component model.  This provides better agreement with the data,
particularly for the Sirius line of sight.  The absorption predicted
by the single-component model may be in conflict with the
absorption limit towards $\epsilon$~Eri near 100 km~s$^{-1}$, but the
conflict is greatly lessened for the multi-component model.  The
greater amount of broad heliosheath absorption for the single-component
model indicates that the simplistic treatment of the plasma in this
model overemphasizes the effects of minority high-temperature
constituents on the plasma velocity distribution.  The effect is to
artificially broaden the distribution, a problem that is then
transmitted to the neutral distributions via charge exchange.

     One very important caveat that must be mentioned is that
the heliospheric models that we have employed here use a spatial grid
that extends only 700~AU from the Sun (see Fig.~1), which may not extend
far enough to capture all the heliospheric absorption in downwind
directions along the lengthy tail of the heliosphere.  Thus, the model
absorption predictions in Figure~6 may end up being lower limits to the
actual amount of absorption that these models would really predict given
a sufficiently large grid.  This problem could potentially worsen
agreement between the models and the data, particularly for the most
downwind directions where the bisector analysis has provided new evidence
for heliosheath absorption.  In the future, we will expand the grid for
the multi-component model up to 10,000 AU, as has already been done for
single-fluid Baranov \& Malama models \citep{vvi03,dba04}.

\section{CONNECTING HELIOSHEATH ABSORPTION WITH LOCAL ENA FLUXES}

     The ASPERA-3 instrument on {\em Mars Express} was designed primarily
to study particles escaping from the Martian atmosphere \citep{sb06}.
However, ASPERA-3 has detected ENAs even when not pointed at Mars, both
en route to Mars and while in orbit around the planet.  \citet{ag06}
and \citet{pw06} have argued that a heliosheath origin for these particles
is more likely than a more local interplanetary source.

    The particle energy spectra observed for the nonplanetary ENAs
are nearly always consistent with the following power law behavior
\citep{ag06}:
\begin{equation}
f(E) \propto \left\{ \begin{array} {c @{\quad\mbox{for}\quad} l}
  E^{-1.6} & E < 0.77~{\rm keV} \\
  E^{-3.3} & E \geq 0.77~{\rm keV}.
  \end{array} \right.
\end{equation}
Particle fluxes are variable, though the nature of the ASPERA-3
observations makes it difficult to distinguish temporal from spatial
variability.  Within ASPERA-3's energy range of $0.2-10$ keV,
logarithmic particle fluxes are in the range $\log F_{A3}=3.7-5.0$ (in
cm$^{-2}$ s$^{-1}$ ster$^{-1}$ units).
If the nonplanetary ENAs observed by ASPERA-3 are indeed
heliosheath neutrals, they then represent the same particle population
that we have been investigating with regards to the Ly$\alpha$
absorption produced by these neutrals.  Thus, we can in principle test
the heliosheath interpretation of the nonplanetary ASPERA-3 ENAs by
seeing if the observed fluxes are consistent with the Ly$\alpha$ data.

     Dotted lines in Figure~7 illustrate the observed energy spectrum
given by equation (1), with the lower and upper lines indicating the
lower and upper range of observed fluxes.  The lower bound of ASPERA-3's
energy range, 0.2~keV, represents a particle velocity of 196 km~s$^{-1}$.
This corresponds with the {\em upper} bound of the $50-200$ km~s$^{-1}$
velocity range where the Ly$\alpha$ data provide the best absorption
constraints (see Fig.~6).  Thus, before we can even hope to use the
ASPERA-3 measurements to predict Ly$\alpha$ absorption, we must
extrapolate the ASPERA-3 energy spectra to lower energies.  There are
three different extrapolations assumed in Figure~7.  The solid line
simply continues the $E^{-1.6}$ power law to lower energies.  The dashed
line assumes a flat spectrum below 0.2~keV, and the dot-dashed line
assumes a drop in flux below 0.2~keV.  The four Figure~7 panels show the
resulting spectra for four different values of $\log F_{A3}$.

     The biggest difficulty in connecting the ASPERA-3 ENA and HST
Ly$\alpha$ data lies in the issue of how to extrapolate a locally
observed particle spectrum along an entire line of sight through the
heliosphere.  If the heliosheath is the source of the ENAs, then for
our purposes the region inside the termination shock (Region~1 in Fig.~1)
is in effect an empty cavity that is being filled by a flux of
heliosheath particles from all directions.  The heliosphere is not
spherically symmetric, so it is likely that fluxes will be larger from
some directions than others.  Nevertheless, assuming that within the
termination shock the particle distributions are isotropic and the same
throughout Region~1 may be a reasonable first-order approximation.
This will probably not be the case within the heliosheath itself.  Thus,
for our purposes here, we will confine our attention only to absorption
by heliosheath neutrals within the termination shock.

     In downwind directions, which are the most appropriate for
heliosheath Ly$\alpha$ absorption, the termination shock distance
is about 200~AU (see Fig.~1).  The inset figures in Figure~7 show the
Ly$\alpha$ absorption produced by heliosheath neutrals within this
distance, assuming that the ENA spectra shown in the figure are isotropic
and apply along the entire 200~AU path length.  The absorption limits
derived for the $\tau$~Cet and $\epsilon$~Eri lines
of sight in \S2.2 (see Fig.~6) are also shown, for comparison with
the absorption predicted by the ENA spectra.

     The two highest flux values explored in Figure~7, $\log F_{A3}=4.55$
and $\log F_{A3}=4.95$, clearly lead to too much Ly$\alpha$ absorption
to be consistent with the limits from the HST data.  A flux of
$\log F_{A3}=4.15$ is only consistent with the Ly$\alpha$ data if fluxes
decrease below 0.2~keV, and even the lowest flux of $\log F_{A3}=3.75$ is
only consistent with the absorption limits if the particle fluxes are flat
or decrease below 0.2~keV.  The situation is actually even worse than this,
given that the absorption predictions in Figure~7 only include absorption
from within the termination shock.  Considering that a downwind line of
sight will have a path length through the heliosheath much longer than
the $\sim 200$~AU path length inside the termination shock, it is clear
that including absorption from the heliosheath could easily increase
the amount of absorption many times above the predictions shown in
Figure~7.  With this in mind, it is very questionable whether {\em any}
of the energy spectra in Figure~7 are consistent with the relative lack
of heliospheric absorption observed in downwind directions.  At the
very least, the spectrum must lie near the lower bound of the
ASPERA-3 range, with fluxes that decrease to lower energies.
It is worth noting that the flux range quoted by \citet{ag06} does not
take into account numerous upper limits when no ENAs were detected
\citep[see Fig.~4 in][]{ag06}, suggesting that perhaps the lower bound
could be decreased further without being truly discrepant with the
ASPERA-3 data.

     Nevertheless, our difficulty in reconciling the ASPERA-3 ENA fluxes
with the HST Ly$\alpha$ spectra is a strong argument against the ASPERA-3
ENAs being heliosheath neutrals.  \citet{ag06} acknowledge other
difficulties with the heliosheath interpretation.  One is that many
heliospheric models predict ENA fluxes at Earth about an order of
magnitude lower than those observed by ASPERA-3 \citep{mg01,pw06}.
The substantial variability seen by ASPERA-3 is also potentially
difficult for a distant heliosheath source to explain, and is more
suggestive of a local interplanetary source.  Identifying the origin of
the ASPERA-3 ENAs is particularly important considering the upcoming
2008 launch of IBEX, which unlike ASPERA-3 is entirely designed and
devoted to studying heliosheath ENAs.  If there is an unexpectedly
strong local interplanetary ENA source that is responsible for the
ASPERA-3 ENAs, IBEX will presumably see it as well and will somehow have
to distinguish between these particles and the true heliosheath ENAs to
accomplish its mission.  It is also worth noting that the ASPERA-4
instrument on board {\em Venus Express} will soon be able to confirm or
refute ASPERA-3's measurements of nonplanetary ENAs.

     Based on Figure~7, we quote a very conservative upper limit of
$F<5000$ cm$^{-2}$ s$^{-1}$ ster$^{-1}$ for the local flux of heliosheath
ENAs in the $0.01-0.2$~keV energy range most applicable to the Ly$\alpha$
absorption data.  The ASPERA-3 and ASPERA-4 instruments are not sensitive
to these low energies, but IBEX will be.  Unless the heliosheath
unexpectedly produces significantly less absorption than the region inside
the termination shock, the actual $0.01-0.2$~keV flux can be expected to
be well below this limit.  More precise constraints on ENA fluxes from the
Ly$\alpha$ data will require the assistance of kinetic heliospheric models
to better connect heliosheath absorption along a downwind line of sight
with the ENA energy spectrum close to the Sun.  Once IBEX is launched it
will be very interesting to see if heliospheric models are capable of
simultaneously reproducing both the local ENA spectrum and the heliosheath
absorption data, particularly the previously detected absorption towards
Sirius and the three new $\theta>160^{\circ}$ detections reported here
for the first time.

\section{SUMMARY}

     The absorption within the Ly$\alpha$ lines of nearby stars
sometimes has a detectable heliospheric contribution, in addition to the
ubiquitous interstellar H~I and D~I absorption.  Most detections are in
upwind directions, where the hydrogen wall dominates the absorption.
However, in this paper we have focused on absorption from the heliosheath,
which is only prominent in downwind directions.
Our findings are as follows:
\begin{description}
\item[1.] In order to search for evidence of previously undetected
  heliosheath absorption, we have measured the bisectors of 28
  stellar Ly$\alpha$ lines reconstructed from the data after correcting
  for the broad ISM absorption (and heliospheric/astrospheric absorption,
  when present).  We find that the stellar Ly$\alpha$ profiles of the
  three most downwind lines of sight ($\chi^{1}$~Ori, HD~28205, and
  HD~28568), all with $\theta > 160^{\circ}$, are significantly
  blueshifted relative to the stellar radial velocities.  This is
  indicative of broad absorption in the red wing of the Ly$\alpha$
  line that was not accounted for in the reconstruction of the stellar
  profile.  This is exactly the signature of heliosheath absorption
  we are looking for, and the most downwind lines of sight are where
  we expect to see it.  Thus, we add these three lines of sight to
  the list of eight previous heliospheric Ly$\alpha$ absorption
  detections, only one of which (Sirius) is a downwind line of sight.
  For the three new detections, the amount of heliospheric absorption
  can be estimated by reanalyzing the Ly$\alpha$ data using a stellar
  profile that is forced to be centered on the stellar rest frame.
\item[2.] The stellar Ly$\alpha$ profiles of the 25 lines of sight with
  $\theta < 160^{\circ}$ are all found to be within $\pm 3$ km~s$^{-1}$
  of the stellar radial velocity.  This represents a conservative
  quantitative estimate of the accuracy of the reconstructed Ly$\alpha$
  profiles.  We can define an upper limit to the red wing of the stellar
  Ly$\alpha$ profile by requiring the bisector to be no more than
  $+3$ km~s$^{-1}$ from either the stellar radial velocity or the bisector
  of the original reconstructed profile, whichever is greatest.  In this
  way, we derive for the first time quantitative upper limits for the
  amount of broad heliosheath absorption that can be present in the red
  wings of Ly$\alpha$ lines without clearly detectable heliosheath
  absorption.
\item[3.] We compare the Ly$\alpha$ absorption predicted by two
  kinetic heliospheric models with downwind Ly$\alpha$ lines observed
  by HST.  One model uses the complex multi-component plasma treatment
  discussed by \citet{ygm06}, while the other uses a
  simple single-fluid approach.  The multi-component model predicts
  significantly less heliosheath absorption than the single-fluid
  model, in better agreement with the data.  This suggests that the
  single-fluid plasma approach, which is the usual assumption in
  heliospheric models, may not reproduce particle distribution functions
  accurately enough for some applications.
\item[4.] Meaningful comparisons between models and the Ly$\alpha$
  absorption data in downwind directions require the use of a model grid
  that extends a sufficiently long distance down the tail of the
  heliosphere to capture all of the heliosheath absorption.  Most
  hydrodynamic models of the heliosphere, including the ones used
  here, do not extend the grid far enough for these purposes.  In the
  future, we hope to perform more precise data/model comparisons using
  sufficiently large grids.
\item[5.] The ASPERA-3 instrument on board {\em Mars Express} has
  detected a substantial flux of nonplanetary ENAs, which have been
  interpreted as being heliosheath neutrals.  We believe that the
  observed ENA fluxes likely predict too much heliosheath
  Ly$\alpha$ absorption in downwind directions to be consistent with
  the HST Ly$\alpha$ spectra.  This represents a strong argument against
  a heliosheath origin for these ENAs, instead suggesting a local
  interplanetary source.
\item[6.] The HST Ly$\alpha$ absorption data can provide useful
  constraints on the heliosheath ENA fluxes that IBEX can expect to
  observe when launched in 2008.  We here quote a very conservative
  upper limit of $F<5000$ cm$^{-2}$ s$^{-1}$ ster$^{-1}$ for the local
  flux of heliosheath ENAs in the $0.01-0.2$~keV energy range most
  applicable to the Ly$\alpha$ absorption data.  We hope to compute more
  stringent constraints in the future using kinetic models of the
  heliosphere to connect particle distributions in the heliosheath with
  those that IBEX will observe locally.
\end{description}

\acknowledgments

We would like to thank Peter Wurz for providing information on the
ASPERA-3 instrument and its measurements.  This work was supported by
NASA grant NNG05GD69G to the University of Colorado.  V. I. was also
supported by RFBR grant 04-02-16559, the ``Dynastia'' Foundation, and
the ``Foundation in Support of Russian Science''.

\clearpage

\begin{deluxetable}{lcccc}
\tabletypesize{\scriptsize}
\tablecaption{Ly$\alpha$ Line Velocities}
\tablecolumns{5}
\tablewidth{0pt}
\tablehead{
  \colhead{Star} & \colhead{$\theta$} & \colhead{$V_{bis}$} &
    \colhead{$V_{rad}$} & \colhead{Ref.} \\
  \colhead{} & \colhead{(deg)} & \colhead{(km s$^{-1}$)} &
    \colhead{(km s$^{-1}$)} & \colhead{}}
\startdata
36 Oph A      &  12.3 &$  0.2$ &$  0.5\pm 0.2$ & 1 \\
70 Oph A      &  23.8 &$ -6.8$ &$ -7.9\pm 0.5$ & 2 \\
HD 165185     &  25.4 &$ 13.5$ &$ 14.9\pm 0.1$ & 3 \\
HD 128987     &  33.3 &$-22.8$ &$-23.0\pm 0.1$ & 3 \\
$\xi$ Boo A   &  46.2 &$  0.2$ &$  1.3\pm 0.1$ & 4 \\
$\alpha$ Cen B&  52.4 &$-20.7$ &$-20.8\pm 0.5$ & 5 \\
61 Vir        &  53.1 &$ -7.5$ &$ -7.9\pm 0.1$ & 4 \\
$\chi$ Her    &  59.6 &$-54.1$ &$-56.1\pm 0.1$ & 4 \\
$\epsilon$ Ind&  69.8 &$-38.9$ &$-38.9\pm 0.3$ & 6 \\
61 Cyg A      &  78.3 &$-65.3$ &$-65.7\pm 0.1$ & 4 \\
HD 116956     &  84.4 &$-12.8$ &$-12.3\pm 0.1$ & 3 \\
HD 97334      &  96.6 &$ -2.0$ &$ -3.7\pm 0.1$ & 4 \\
EV Lac        &  98.2 &$  3.0$ &$  0.4\pm 0.1$ & 4 \\
$\zeta$ Dor   & 107.7 &$ -1.4$ &$ -0.7\pm 0.1$ & 7 \\
HD 166        & 111.0 &$ -7.5$ &$ -6.5\pm 0.1$ & 4 \\
DK UMa        & 111.5 &$-27.6$ &$-27.3\pm 0.1$ & 8 \\
$\beta$ Cas   & 111.8 &$ 13.3$ &$ 11.3\pm 0.5$ & 9 \\
HD 73350      & 122.1 &$ 33.8$ &$ 35.4\pm 0.1$ & 4 \\
$\tau$ Cet    & 122.9 &$-18.2$ &$-16.6\pm 0.1$ & 4 \\
HD 43162      & 137.3 &$ 22.1$ &$ 21.7\pm 0.1$ & 3 \\
HD 37394      & 140.3 &$  1.1$ &$  1.2\pm 0.1$ & 4 \\
$\epsilon$ Eri& 147.4 &$ 17.9$ &$ 16.3\pm 0.1$ & 4 \\
$\delta$ Eri  & 148.8 &$ -8.8$ &$ -6.3\pm 0.1$ & 4 \\
$\kappa$ Cet  & 152.5 &$ 16.4$ &$ 19.0\pm 0.1$ & 4 \\
40 Eri A      & 154.7 &$-42.4$ &$-42.3\pm 0.1$ & 4 \\
$\chi^1$ Ori  & 166.3 &$-21.1$ &$-15.3\pm 0.5$ & 10,11 \\
HD 28205      & 172.0 &$ 33.2$ &$ 39.3\pm 0.2$ & 12 \\
HD 28568      & 173.3 &$ 30.3$ &$ 40.9\pm 1.3$ & 12 \\
\enddata
\tablerefs{(1) Beavers \& Eitter 1986. (2) Tokovinin \& Smekhov 2002.
  (3) Gaidos et al.\ 2000. (4) Nidever et al.\ 2002.
  (5) Pourbaix et al.\ 2002. (6) Buscombe \& Kennedy 1968.
  (7) Murdoch et al.\ 1993. (8) Duquennoy et al.\ 1991.
  (9) Duflot et al.\ 1995. (10) Gullberg \& Lindegren 2002.
  (11) Irwin et al.\ 1992. (12) Perryman et al.\ 1998.}
\end{deluxetable}

\clearpage

\begin{figure}
\plotone{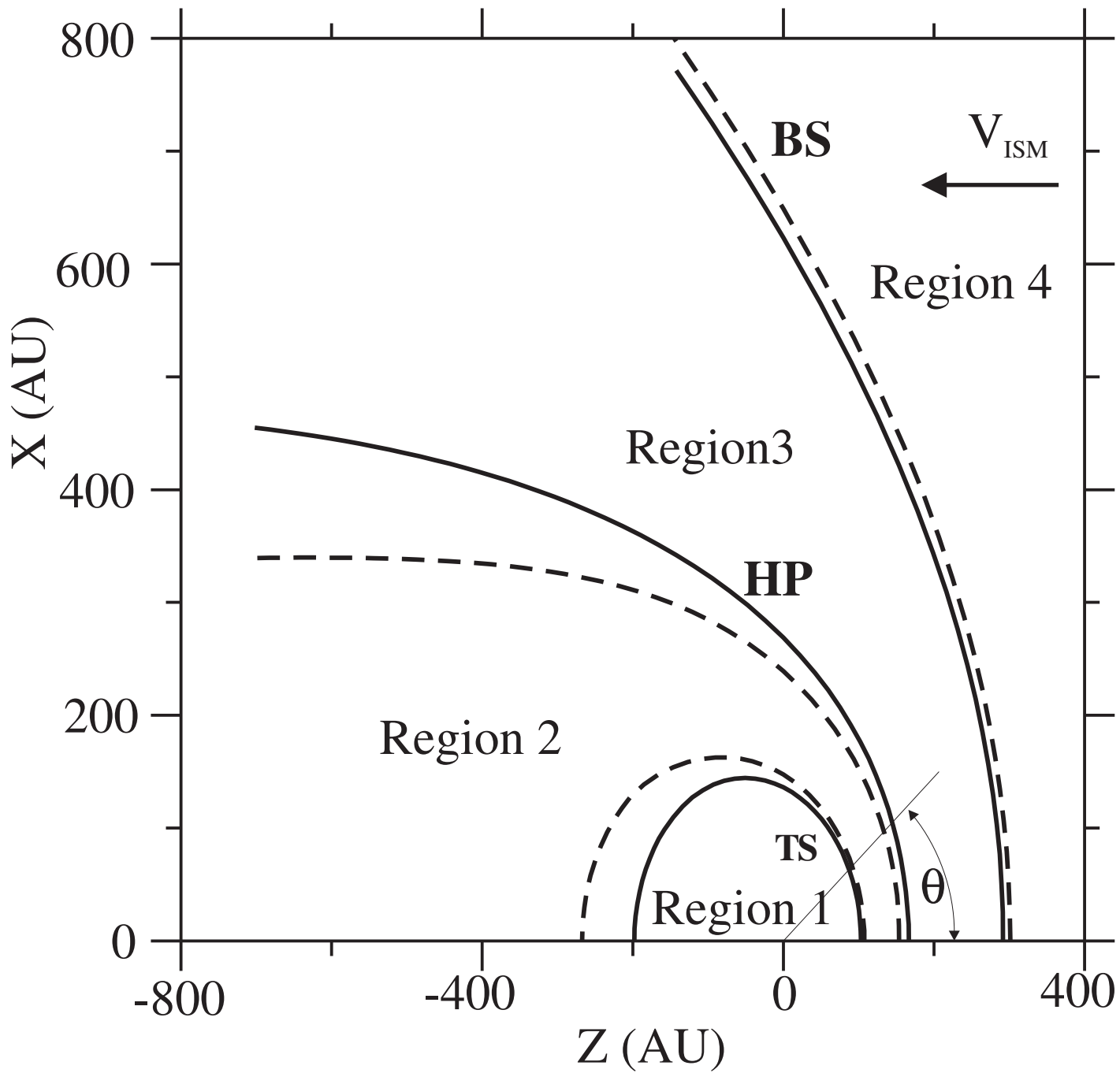}
\caption{The locations of the termination shock (TS), heliopause (HP),
  and bow shock (BS) according to the single-fluid (solid
  lines) and multi-component (dashed lines) plasma models used in the
  text (see \S2.1 and \S3).  The Sun is at the origin.  The ISM flow
  direction is indicated by the V$_{ISM}$ vector, and the angle with
  respect to the upwind direction of this flow angle, $\theta$, is
  also illustrated.}
\end{figure}

\begin{figure}
\plotone{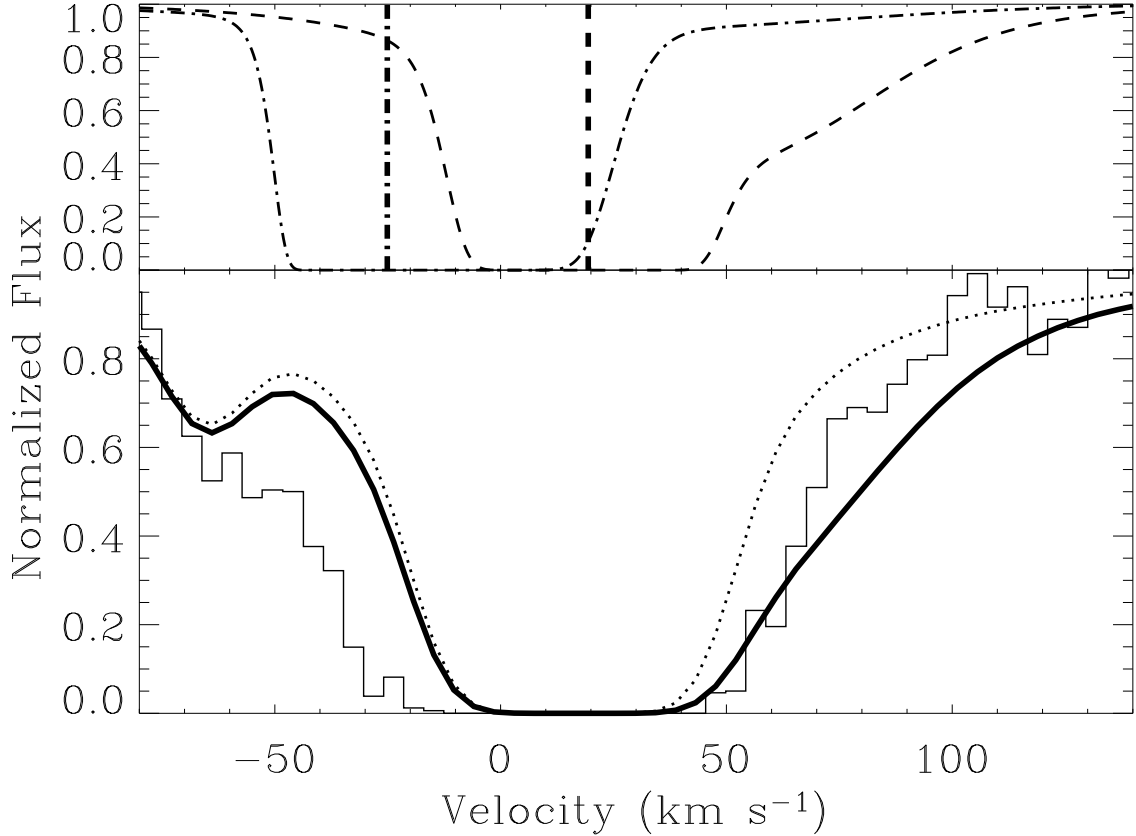}
\caption{The top panel compares heliospheric H~I Ly$\alpha$ absorption
  profiles plotted on a heliocentric velocity scale predicted for the
  upwind 36~Oph (dot-dashed line) and downwind Sirius (dashed line) lines
  of sight, based on a hydrodynamic model of the heliosphere (see text).
  The thick vertical lines indicate the projected ISM flow speed
  in these two directions.  The bottom panel shows an HST spectrum of
  the Ly$\alpha$ absorption line observed towards Sirius (histogram).
  The dotted line is the ISM absorption alone for this line of sight.
  The broad absorption centered at +20 km~s$^{-1}$ is interstellar H~I
  absorption and the weak absorption at about $-65$ km~s$^{-1}$ is from
  interstellar deuterium.  The thick solid line is the predicted
  heliospheric absorption from the upper panel (dashed line) combined with
  the ISM absorption.}
\end{figure}

\begin{figure}
\plotone{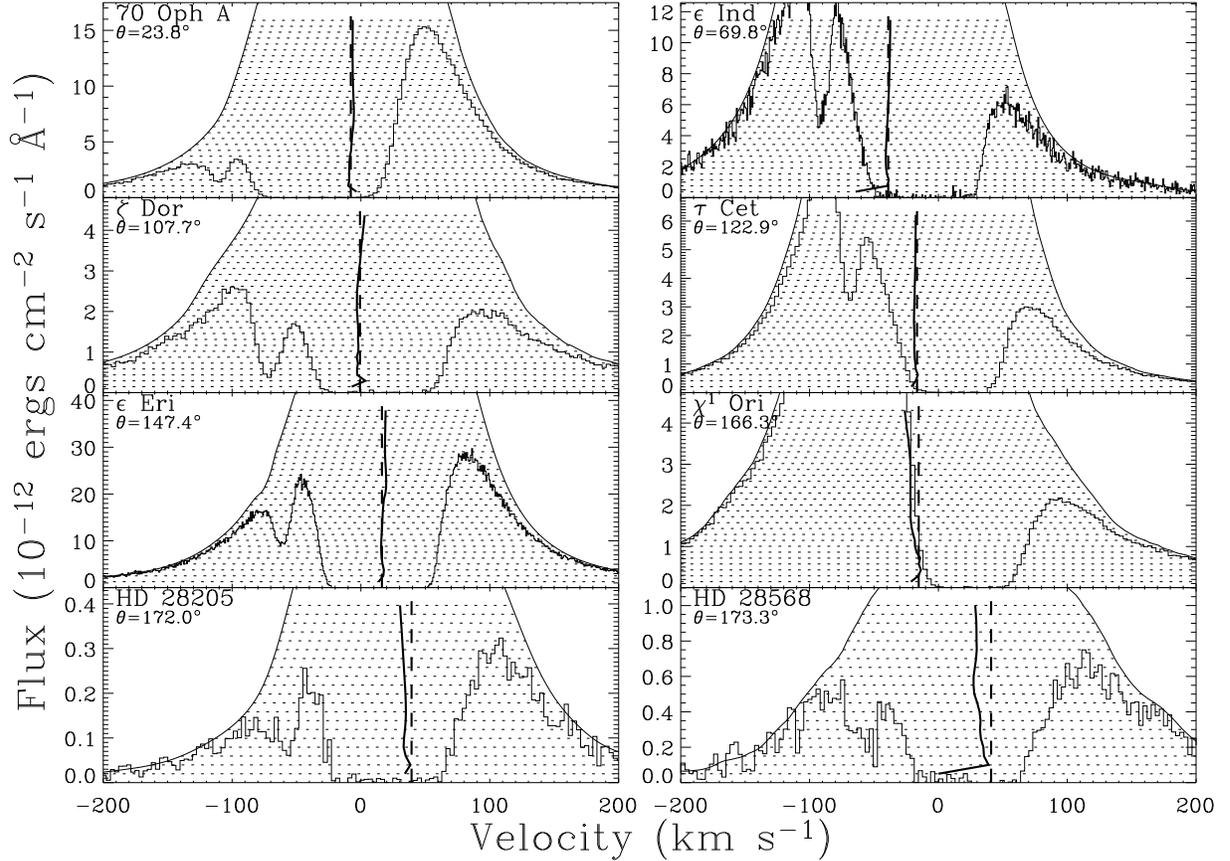}
\caption{Ly$\alpha$ line bisector measurements for various stars.  The
  histograms are the HST Ly$\alpha$ spectra.  The broad absorption
  dominating the observed line profiles is H~I absorption, and the weak
  absorption about $-82$ km~s$^{-1}$ from the center of the H~I
  absorption is from interstellar deuterium.  The solid lines
  above the data are the reconstructed stellar Ly$\alpha$ profiles.
  The dotted lines indicate the flux levels where the bisectors of
  the stellar profiles are measured, which are the thick near-vertical
  solid lines.  The vertical dashed lines are the stellar radial
  velocities.  Note that for the three most downwind lines of sight
  ($\chi^{1}$~Ori, HD~28205, and HD~28568) the bisector is blueshifted
  relative to the stellar rest frame, suggesting the presence of
  heliosheath absorption in the red wing of the line.}
\end{figure}

\begin{figure}
\plotone{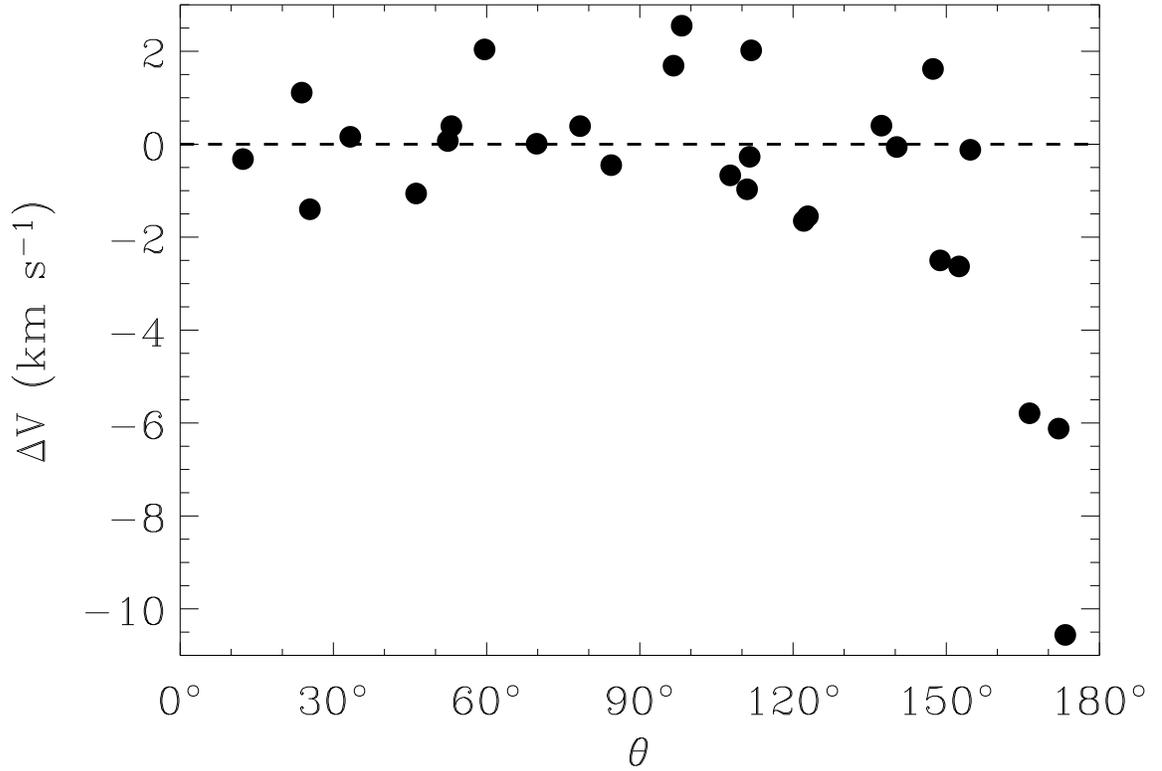}
\caption{Velocity differences between the Ly$\alpha$ line bisector
  and the stellar rest frame ($\Delta V\equiv V_{bis}-V_{rad}$) plotted
  as a function of $\theta$, the line-of-sight orientation with respect
  to the ISM flow vector.  There is evidence that Ly$\alpha$ profiles
  reconstructed for the most downwind lines of sight
  ($\theta>160^{\circ}$) are systematically blueshifted with respect to
  the stellar rest frame, suggesting the presence of heliosheath
  absorption.}
\end{figure}

\begin{figure}
\plotone{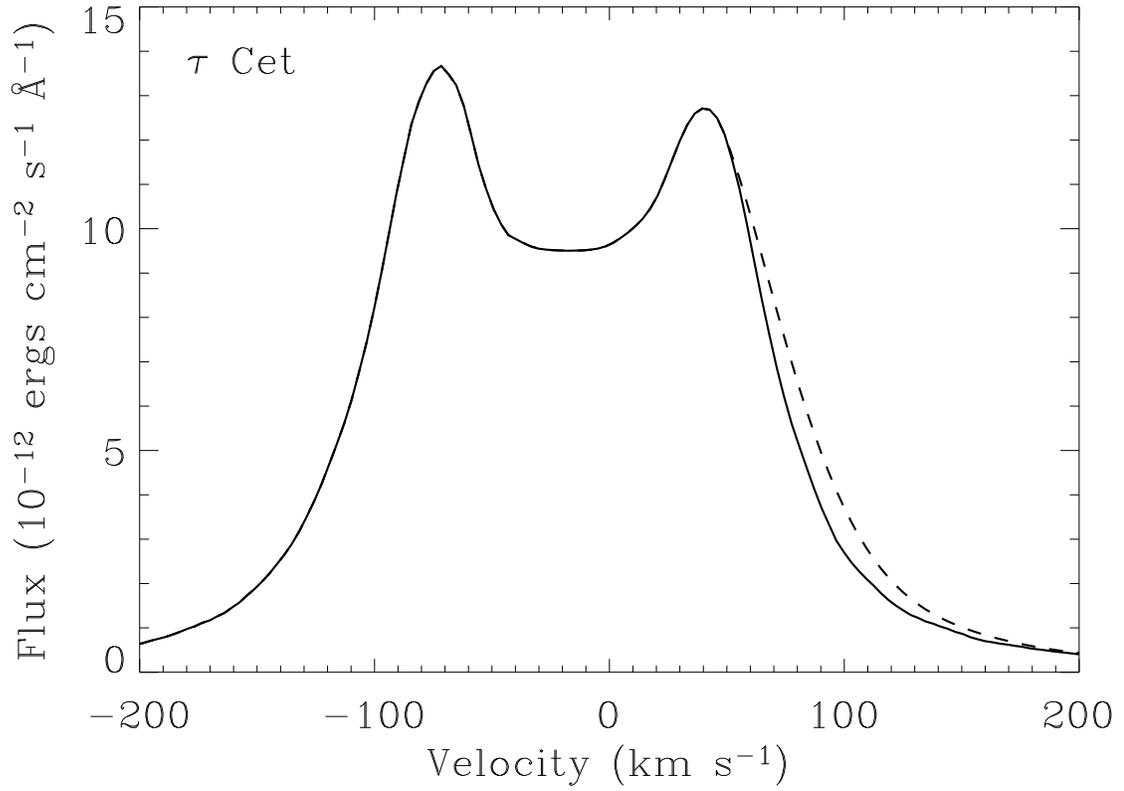}
\caption{The reconstructed stellar Ly$\alpha$ profile for
  $\tau$~Cet (solid line), and an upper limit for the red wing of the
  profile (dashed line) derived by requiring that the profile be within
  3 km~s$^{-1}$ of the stellar radial velocity (see text).}
\end{figure}

\begin{figure}
\plotone{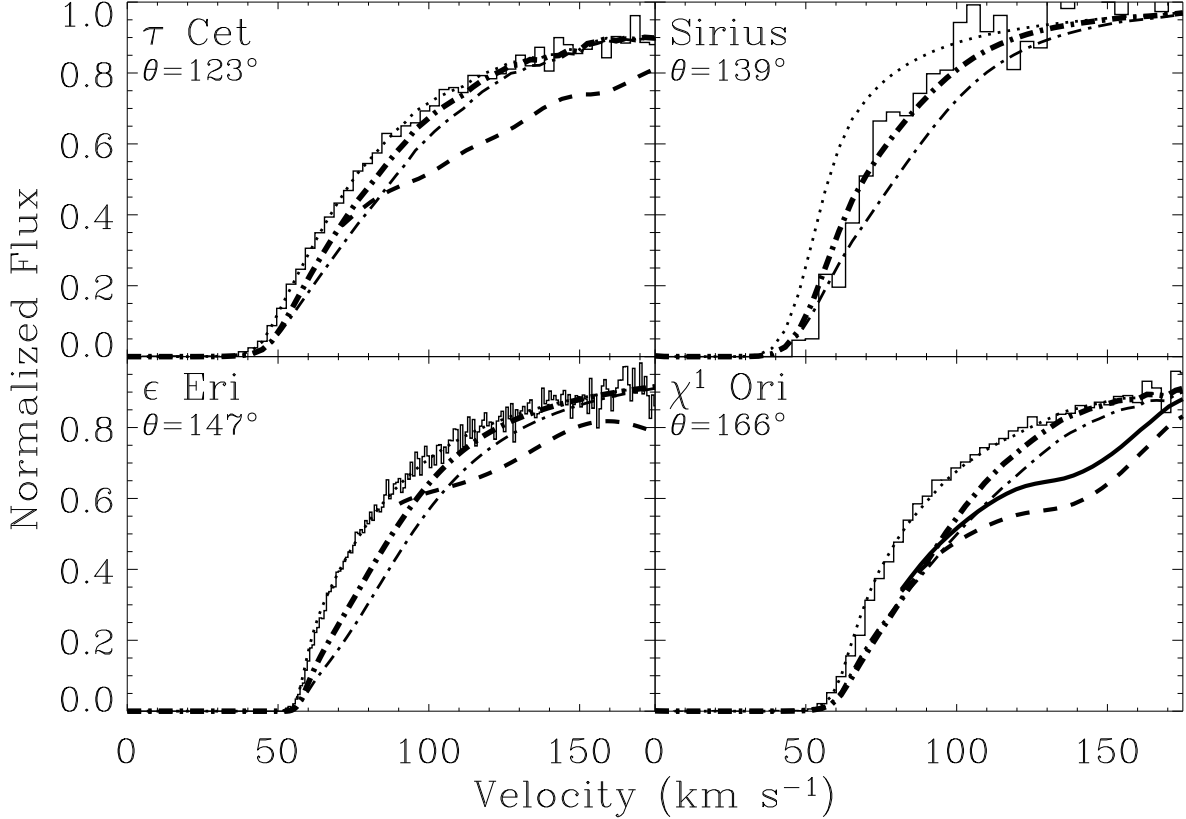}
\caption{Normalized Ly$\alpha$ spectra in downwind directions, focusing
  on the red side of the Ly$\alpha$ absorption line where the heliospheric
  absorption resides.  The dotted line is the ISM absorption alone, which
  can account for all the observed absorption, except for Sirius.  The
  thin and thick dot-dashed lines are the predicted heliospheric
  Ly$\alpha$ absorption of models with a single plasma fluid component and
  a multi-component plasma treatment, respectively (see text).  The dashed
  lines (when present) are the maximum heliospheric absorption that is
  allowable by the data, based on upper limit stellar Ly$\alpha$ profiles
  derived as described in the text.  The models can be considered
  consistent with the data if their predicted absorption lies above these
  dashed lines.  For $\chi^{1}$~Ori, there is evidence from a blueshift of
  the original reconstructed Ly$\alpha$ line that there is extended
  heliosheath absorption present, and the thick solid line is
  the amount of absorption that should be observed if the real stellar
  Ly$\alpha$ line is centered on the stellar rest frame.}
\end{figure}

\begin{figure}
\plotone{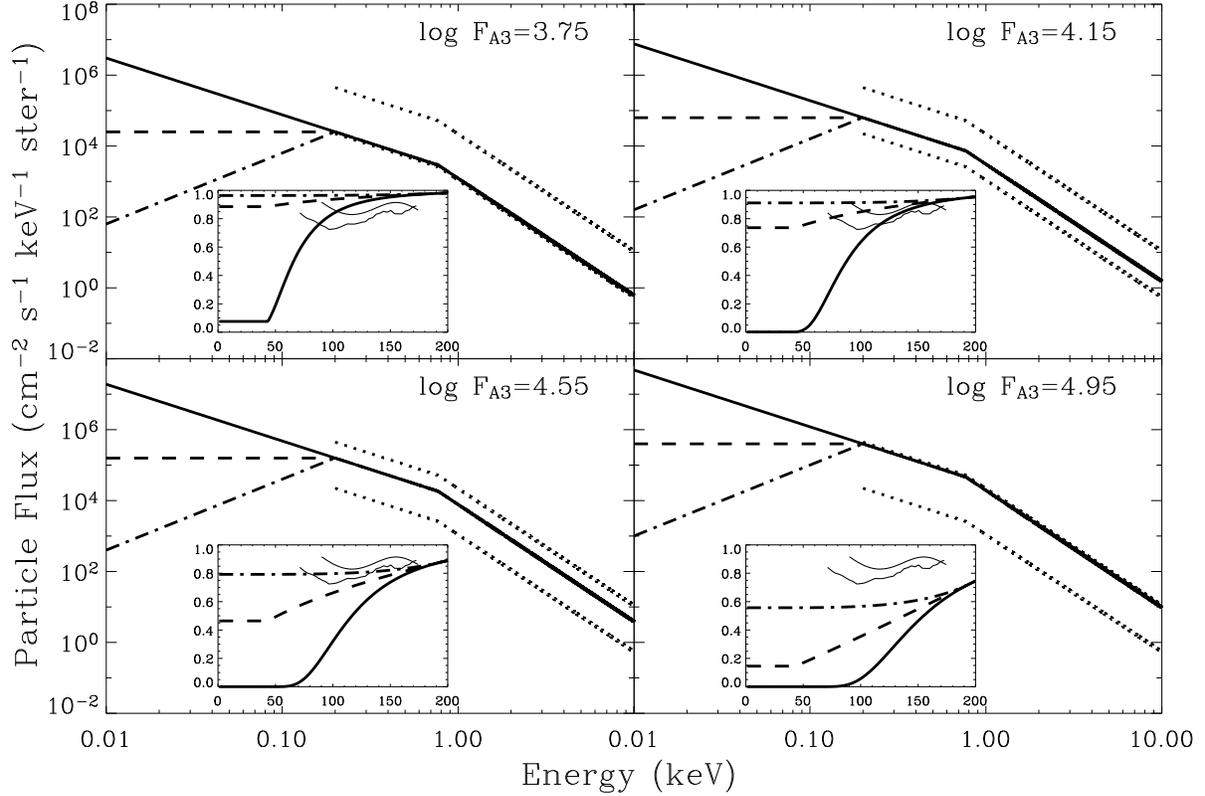}
\caption{The dotted lines in each panel encompass the range of ENA
  spectra observed by ASPERA-3, corresponding to different $0.2-10$~keV
  particle fluxes (in cm$^{-2}$ s$^{-1}$ ster$^{-1}$ units) of
  $\log F_{A3}=3.7-5.0$.  Assuming different fluxes within this range,
  each panel shows three different spectra, which use different
  extrapolations to low energies.  The inset figures show the H~I
  Ly$\alpha$ absorption from within the termination shock predicted by
  these spectra, plotted on a heliocentric velocity scale (in km~s$^{-1}$).
  The two thin solid lines that stretch from $\sim 70$ to 175 km~s$^{-1}$
  are the absorption limits derived for $\tau$~Cet and $\epsilon$~Eri
  based on the Ly$\alpha$ bisector analysis (see \S2.2 and Fig.~6).
  The absorption predicted by the energy spectra must lie above these
  limits to be consistent with the Ly$\alpha$ data.}
\end{figure}

\end{document}